\documentclass{aa}
\usepackage{graphics}
\usepackage{times}
\def\cm{\,{\rm cm}}
\def\cm3{\,{\rm cm^{-3}}}

\def\ion#1#2{\ifmmode \mbox{{\rm #1}}\,\mbox{{\sc #2}} % generic ion
        \else {\rm #1}$\,${\sc #2}
        \fi}

\def\kms{\,{\rm {km\,s^{-1}}}}
\def\kkms{\,{\rm {K\,km\,s^{-1}}}}

\def\hmsd#1h#2m#3.#4s{                  % hms format with decimal point (RA)
                      \relax
                      \ifmmode #1^{\rm h}\,#2^{\rm m}\,#3^{\rm s}
                               \hskip-0.39em.\hskip0.08em#4
                      \else \hbox{$#1^{\rm h}\,#2^{\rm m}\,#3^{\rm s}
                            \hskip-0.39em.\hskip0.08em#4$}
                      \fi
                     }
\def\dms#1d#2m#3s{                      % dms format (for Dec)
                  \relax
                  \ifmmode #1^\circ\,#2^{\prime}\,#3^{\prime\prime}
                  \else \hbox{$#1^\circ\,#2^{\prime}\,#3^{\prime\prime}$}
                  \fi
                 }
\def\degd#1.#2{                         % degrees over decimal point
               \ifmmode {#1^{\hskip 0.05em\circ}\hskip-0.42em.\hskip0.08em#2}
               \else {#1$^{\hskip 0.05em\circ}\hskip-0.42em.\hskip0.08em$#2}
               \fi
              }
\def\mind#1.#2{                         % minutes over decimal point
               \ifmmode {#1^{\hskip 0.05em\prime}\hskip-0.35em.\hskip0.05em#2}
               \else {#1$^{\hskip 0.05em\prime}\hskip-0.35em.\hskip0.05em$#2}
               \fi
              }
\def\secd#1.#2{                         % seconds over decimal point
               \ifmmode {#1^{\prime\prime}\hskip-0.46em.\hskip0.12em#2}
               \else {#1$^{\prime\prime}\hskip-0.46em.\hskip0.12em$#2}
               \fi
              }
\def\mg{\relax                          % magnitudes symbol
        \ifmmode ^{\rm m}
        \else $^{\rm m}$
        \fi
       }
\def\mgd#1.#2{                          % magnitudes over decimal point
              \relax
              \ifmmode #1^{\rm m}
                       \hskip-0.55em.\hskip0.22em#2
              \else \hbox{#1$^{\rm m}
                    \hskip-0.55em.\hskip0.22em$#2}
              \fi
             }
\def\unitspace{\,}                      % space to be used with units

\def\un#1{\ifmmode \unitspace{\rm #1} % generic unit
          \else $\unitspace$#1
          \fi}
\def\pun#1#2{\ifmmode \unitspace\mbox{\rm #1}^{#2} % generic unit with a power
             \else $\unitspace$#1$^{#2}$
             \fi}
\def\mum{\ifmmode \unitspace\mu\mbox{\rm m} % micron
         \else $\unitspace\mu$m
         \fi}
\def\aua{{A\&A} }

\def\apj{{ApJ} }
\def\aj{{AJ} }
\def\apjs{{ApJS} }
\def\apjl{{ApJ} }

\def\pasj{{PASJ} }

\begin{document}
 
\thesaurus{03         % A&A Section 3: Extragalactic
          (11.01.2;
           11.09.1 NGC\,891; 
           11.09.4; 
           11.14.1;
           13.09.1)
          }

\title{Submillimetre maps of the edge-on galaxy NGC\,891}
 
\author{F.P.\ Israel\inst{1}
 \and   P.P.\ van der Werf\inst{1}
 \and	R.P.J.\ Tilanus\inst{2,3}
       }
 
\offprints{F.P.\ Israel}
\mail{israel@strw.leidenuniv.nl} 

\institute{$^1$ Sterrewacht Leiden, P.O.\ Box 9513, NL - 2300~RA Leiden,
             The Netherlands\\
$^2$ Joint Astronomy Centre, 660 N.~A'ohoku Pl., Hilo, Hawaii, 96720, USA\\
$^3$ Netherlands Foundation for Research in Astronomy, P.O. Box 2, 7990 AA 
     Dwingeloo, The Netherlands
}
 
\date{Received ... / Accepted ...}
 
\maketitle

\begin{abstract}
Broadband continuum images of the edge-on galaxy NGC\,891 at 850\,$\mu$m 
and 450\,$\mu$m are presented. These images are very similar
to the 1300$\mu$m and CO line images. With respect to 
the 850$\mu$m continuum, CO is strongest in the socalled molecular ring',
and weakest at the largest radii sampled. Inside the molecular ring,
the CO/850$\mu$m ratio is somewhat less than in the ring, but higher
than in the remainder of the disk. The integrated infrared emission
from NGC~891 is dominated by small particles shortwards of 100$\mu$m.
Longwards of 100$\mu$m, the emission can be equally well fitted by
a single population of large dust grains at 21 K, or a bimodal
population of grains at temperatures of 18 K and 27 K. The circumnuclear
disk is at a temperature of at least 50 K, and probably much higher.


\end{abstract}
 
\section{Introduction}

NGC\,891 is a bright, non-interacting spiral galaxy (SA(s)b: De Vaucouleurs 
et al.\ 1976). It is very nearly edge-on ($i > 88^{\circ}$, $PA = 23^{\circ}$;
Sofue $\&$ Nakai 1993). NGC~891 is one of the major members of the 
NGC\,1023 group and its distance is estimated to be 9.5\,Mpc for 
$H_{\rm 0}=75\kms\,{\rm Mpc}^{-1}$, making for a scale of 22$''$/kpc or 
46\,pc/$''$. Many authors have remarked on the apparent similarity between 
NGC~891 and the Milky Way. The interstellar medium in the disk of NGC~891 
has been the subject of several studies, notably radio continuum (Allen et 
al. 1978; Sukumar $\&$ Allen 1991), HI (Sancisi $\&$ Allen 1979; Rupen 1991), 
dust (Howk $\&$ Savage 1997); CO (Garc\'ia-Burillo et al. 1992; Scoville et 
al. 1993; Sofue $\&$ Nakai 1993; Garc\'ia-Burillo $\&$ Gu\'elin 1995; 
Sakamoto et al. 1997) and CI (Gerin $\&$ Phillips 1997). Far-infrared dust 
emission was observed by IRAS (see Wainscoat et al. 1987; Rice et al. 1988), 
and at $\lambda$ = 
1.3 mm by Gu\'elin et al. 1993. In this paper we present further observations 
of the submillimeter dust emission from NGC~891 with the new SCUBA 
submillimeter continuum array detector.

\section{Observations}

The SCUBA 850 and 450 $\mu$m images were obtained in 1997 September 
on the JCMT \footnote{The James Clerk Maxwell Telescope is operated by the
Joint Astronomy Centre on behalf of the Particle Physics and Astronomy 
Research Council of the United Kingdom, the Netherlands Organisation for
Scientific Research and the National Research Council of Canada}. 
SCUBA employs two arrays of cooled bolometers, each
covering a field of about $\mind 2.3$. The respective filters used 
have central frequencies of 347 GHz and 677 GHz respectively; both 
filters have a bandwidth of 30 GHz. We observed three fields on
NGC~891 (center, northeast and southwest) in socalled `jiggle-mode'
in order to cover the whole galaxy. Chopping distance was 2$'$
perpendicular to the major axis of the galaxy. Total integration time 
for the three fields was 128 minutes. Reduction was performed in the
standard manner. Photometric calibration was achieved by skydip
analysis and mapping of the standard star CRL 618. More complete 
details of the instrument as well as observing and reduction methods 
can be found in Holland et al. (1998) and Jenness et al. (1998), as 
well as on the world-wide web (www.jach.hawaii.edu). 
The resulting images are shown in Figure 1. Especially in the 450$\mu$m
image, residual noise is seen at the SCUBA field edges; this should
be discounted.

\begin{figure*}[tph]
\begin{center}
\resizebox{!}{18cm}{\includegraphics{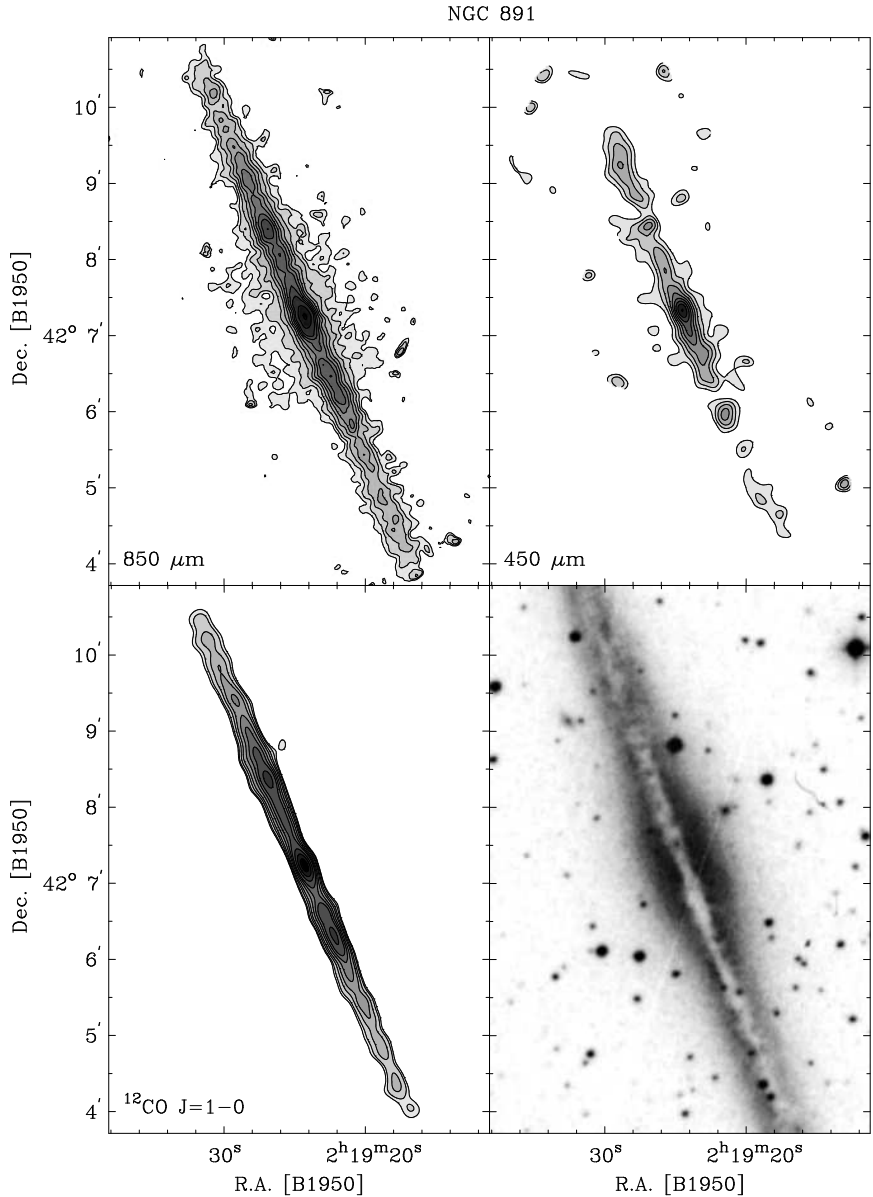}}
\end{center}
\vspace{-1cm}
\caption[]{
Submillimeter continuum images at 13$''$ resolution of NGC\,891 at 
850$\mu$m (top left) and 450$\mu$m (top right). Contour levels are 15, 
30, 50, 75, 100, 125, 150, 175, 200, 250 and 300 mJy/beam (850$\mu$m), 
500, 750, 1000, 1250, 1500, 1750, 2000, 2250 and 2500 mJy/beam (450$\mu$m).
For comparison, the interferometric $J$=1--0 $^{12}$CO map obtained by 
Scoville et al. (1993) is included at bottom left, and an optical image 
at bottom right.
}
\label{fig.SCUBA}
\end{figure*}

\section{Results}

\begin{figure*}[t]
\begin{center}
\resizebox{!}{14cm}{\includegraphics{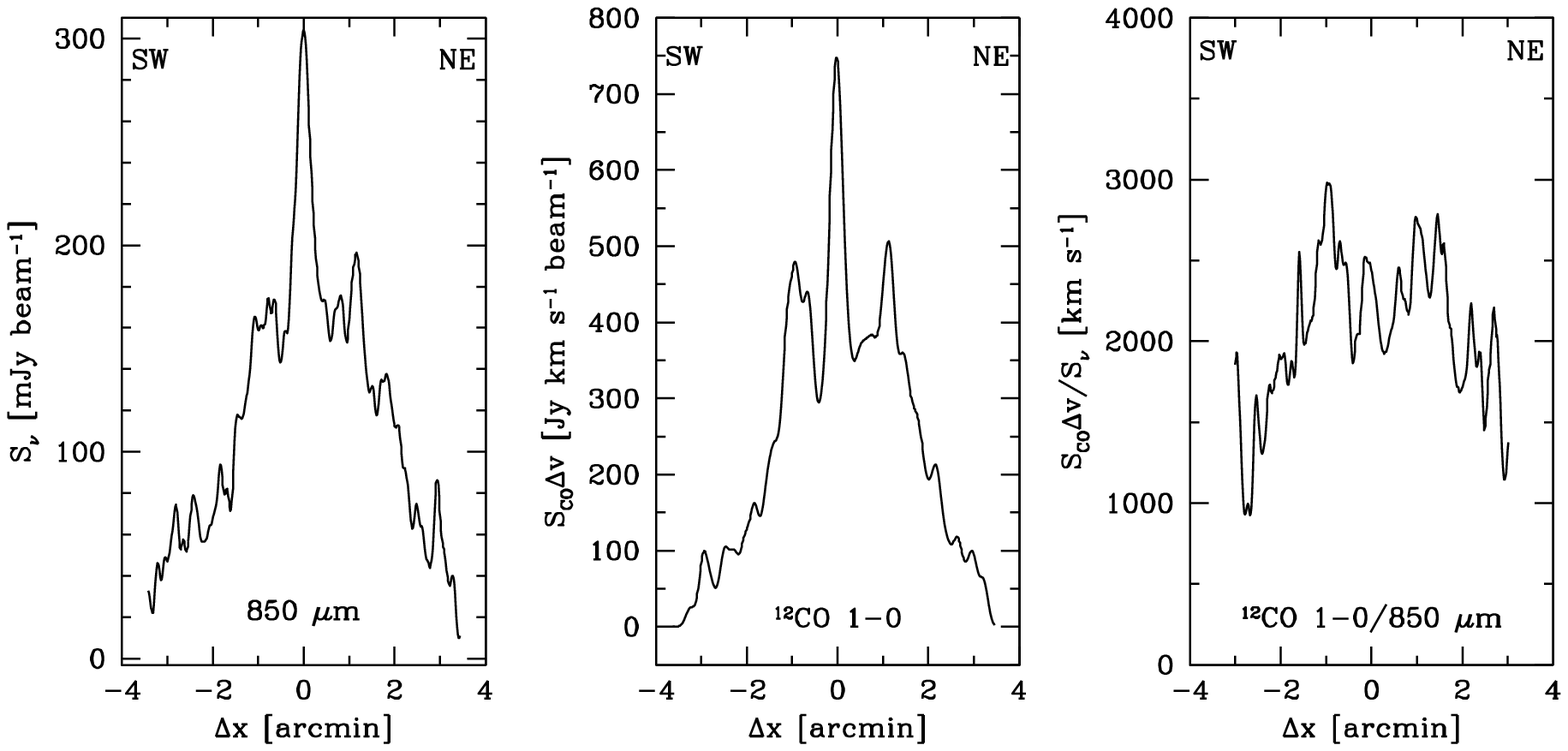}}
\end{center}
\vspace{-7cm}
\caption[]{Radial distribution of 850\,$\mu$m (left), $J$=1--0 $^{12}$CO 
(center) emission in NGC\,891 and the ratio of observed CO/850$\mu$m 
emission (right). Profiles show emission at 13$''$ resolution along the 
major axis.
}
\label{fig.cuts}
\end{figure*}

\subsection{Total submillimetre spectrum and dust content}

Gu\'elin et al. (1993) have fitted two dust components at 15 K and 30
K to their 1.3 mm flux and the IRAS 100$\mu$m flux. From our maps, we 
have determined integrated flux-densities $S_{850}$ = 4.8$\pm$0.6 Jy and 
$S_{450} = 39\pm7$ Jy. The observations through the 850$\mu$m filter 
include emission from the $J$=3--2 $^{12}$CO line. In similar beams, 
the first three CO transitions occur in intensity ratios of 1:0.75:0.4
(see Garc\'ia-Burillo et al. 1992; Gerin $\&$ Phillips 1997). From
this, we estimate an integrated $J$=3--2 CO contribution of 0.22 Jy 
to the measured 850$\mu$m flux-density. Corrected for this contribution,
the dust continuum emission becomes $S_{850} \approx$ 4.6 Jy. The 
$J$=6--5 $^{12}$CO transition is at the edge of the 450$\mu$m filter; 
we expect its contribution to the observed flux to be negligible.
With $S_{1300}$ = 0.73 Jy (Gu\'elin et al. 1993), and the IRAS flux 
densities $S_{12}$ = 6 Jy, $S_{25}$ = 7 Jy, $S_{60}$ = 80 Jy and $S_{100}$ 
= 196 Jy (Wainscoat et al. 1987; Rice et al. 1988; Young et al. 1989), 
the infrared/submillimeter spectrum of NGC~891 is now well-determined.  

Assuming a dust emissivity proportional to $\lambda^{-2}$, we have
fitted these flux-densities with various dust temperatures.

(i). Longwards of 100$\mu$m, a good fit is obtained with a {\it single} 
dust temperature $T_{\rm d} \approx$ 21 K. However, this fit leaves the 
mid-infrared fluxes and most of the 60$\mu$m flux to be explained. 
The interstellar dust models by D\'esert, Boulanger $\&$ Puget (1990) 
suggest that virtually all of the 12$\mu$m flux and about 50$\%$ of the 
25$\mu$m flux is attributable to PAH particles. This implies a ratio 
$F_{25}/F_{60}$ = 0.1 for the remaining very-small grain (VSG) 
component which suggests an average VSG size somewhat larger than in 
the Solar Neighbourhood, i.e. about 10 nm instead of 7 nm (see D\'esert 
et al. 1993). We have used equation (2) from Gu\'elin et al. (1993) to 
calculate the total hydrogen mass associated with the radiating dust 
(assuming $b(Z/Z_{\odot}$ = 1.6) and find $M_{\rm H} = 3 \times 10^{9}$ 
M$_{\odot}$.

(ii). Next, we assume that the 12 - 1300 $\mu$m spectrum is described by
the dust model of D\'esert et al. (1990), with big-grain dust radiating
at two different temperatures. With dust temperatures $T_{\rm wd}$  = 
28$\pm$2 K and $T_{\rm cd}$ = 18$\pm$1 K, close to the values assumed 
by Gu\'elin et al. (1993), calculated fluxes reproduce the observed values
to better than 20$\%$ over the full wavelength range. At 850$\mu$m, we
find flux-densities of 1.2 Jy (VSG), 0.55$\pm$0.15 Jy (warm dust) and 
3.1$\pm$0.2 Jy (cold dust). From this, we obtain $M_{\rm H}(warm) = 
3.3\pm1.1 \times 10^{8}$ M$_{\odot}$ and $M_{\rm H}(cold) = 3.6\pm0.4 
\times 10^{9}$ M$_{\odot}$, or a total mass $M_{rm H}(total) = 3.9\pm0.3
\times 10^{9}$ M$_{\odot}$, slightly less than that estimated by Gu\'elin 
et al. (1993). The ratio $M_{\rm H}(warm)/M_{\rm H}(cold) = 0.1\pm0.04$ is 
higher than surmised by Gu\'elin et al. (1993).

Because the dust emission is a strong function of temperature, the 
temperatures are relatively well-determined, depending on the choice of
one or two temperature components. As is clear from the above, the choice
of temperature model has relatrively little effect on the derived mass. 
Following Gu\'elin et al (1993), we adopt $M(HI) = 2.5 \times 10^{9}$ 
M$_{\odot}$ so that $M(H_{2}) = 1.4 \times 10^{9}$ M$_{\odot}$. this 
yields a conversion factor $X = N(H_{2})/I_{CO} \approx 7 \times 10^{19}$ 
mol cm$^{-2}$ $(\kkms)^{-1}$, somewhat less than 
found by Gu\'elin et al. (1993) and Gerin $\&$ Phillips (1997) and 
substantially less than the canonical value. Assuming $M(HI)_{\rm tot} = 
3.7 \times 10^{9}$ (Allen $\&$ Sancisi, 1979; Rupen 1991) and all H$_{2}$ to
be sampled, the overall ratio of molecular to atomic hydrogen is 
0.38, the molecular gas fraction is 0.20 and within $R=16.5$ kpc, the total 
gas fraction, including helium, is 5\%. The first two values are 
similar to those found for the Milky Way and for dwarf irregular galaxies 
(Israel 1997). 

\subsection{Radial distributions}

The structure of CO and dust continuum emission along the major axis can 
be roughly described by three components: the very compact central source 
at (a circumnuclear disk -- Garc\'ia-Burillo et al. 1992; Scoville 
et al. 1993; Sofue $\&$ Nakai 1993), a `molecular ring' between $R = 
40''$ and $R = 120''$ (2 -- 6 kpc) and the extended disk traceable out 
to $R = 200''$ (9 kpc). Given the edge-on configuration of NGC~891, the 
various peaks in the emission are most likely spiral arms seen tangentially. 
This includes the bright inner peaks which may well be due to bright spiral 
arms encircling the bulge rather than a ring. 

As Figure 1 shows, the 850$\mu$m (and 450$\mu$m) continuum image is 
virtually identical to the $J$=1--0 CO image. This is in line with the 
close resemblance of the 1.3 mm to the $J$=2--1 $^{12}$CO distribution 
(Gu\'elin et al. 1993) and indeed to the $^{3}P_{1} - ^{3}P_{0}$ CI 
distribution (Gerin $\&$ Phillips 1997) and the 6 cm continuum emission
(see Allen et al. 1978). This is further illustrated by the radial 
distributions shown in Figure 2. Figure 2c shows $J$=1--0 CO/850$\mu$m 
emission ratios of 1500 for the central peak and 2000 for the molecular ring 
peaks (or inner spiral arm tangential points), dropping to 1300 for the
inner region inside the `ring'. 

Although decreasing signal-to-noise ratios introduce increasingly large 
relative errors in the outer disk ratios, there appears to be a steady 
decrease in this ratio, in the outer disk dropping to 70$\%$ of the value
at $R = \pm2'$, just outside the molecular ring. The run of the 450/850/1300 
$\mu$m intensities with radius provides no support for significant dust 
temperature gradients. We propose that the high ratio of the molecular 
ring primarily reflects a higher CO surface filling factor. If the inner 
spiral arms contain a rich population of dense CO clouds, this will quite 
naturally produce a relatively high CO filling factor (hence a high ratio 
in Figure 2c) at the arm tangential points. High line-of-sight CO optical 
depths are suggested by the $^{12}$CO/$^{13}$CO ratio of 2.5 for the ring 
(Sakamoto et al. 1997).

The decrease of the CO/850$\mu$m ratio with increasing distance to the center 
parallels the radial drop in CO emissivity shown by Scoville et al. (1993)
and the even steeper decrease of $^{13}$CO emission (Sakamoto et al. 1997),
resulting in a $^{12}$CO/$^{13}$CO gradient increasing from 5 just outside 
the molecular ring to about 15 in the outer disk. The latter interpret this
as a steep radial decrease of the dense gas fraction, so that we may expect
the effects of photo-dissociation processes to increase radially. This appears
to be confirmed by the observed major axis distributions of [CI] (Gerin 
$\&$ Phillips 1997) and [CII] (Madden et al. 1994). 

The central region is quite different. Virtually all major axis profiles
exhibit a pronounced dip inside the molecular ring, onto which the compact
emission from the circumnuclear disk is superposed. This suggests that the
region {\it inside the molecular ring}, i.e. the region within $R$ = 2 kpc
corresponding to the bulge of NGC~891, contains {\it little}  
interstellar material except for the circumnuclear disk of radius 600 pc.
The emission observed towards the center mostly represents material 
in the molecular ring with a smaller contribution from the outer disk.
The cavity appears to be filled by very hot ($T = 3.6 \times 10^{6}$ K)
X-ray-emitting gas, extending into the molecular ring (Bregman $\&$ Pildis
1994). West of the nucleus, an X-ray jet feature is seen. This, and
the presence of a nuclear disk suggests that NGC~891 has a mildly active
nucleus. We have attempted to separate the nuclear continuum emission peak
from the more extended emission. We find fluxes $S_{1300}$ = 55 mJy,
$S_{850}$ = 230 mJy and $S_{450}$ = 2200 mJy. Corrected for
CO line contributions (Garc\'ia-Burillo et al. 1992; Gerin $\&$ Phillips 
1997), these reduce to 40, 190 and 2200 mJy respectively. Although these
fluxes are relatively uncertain, they define a spectrum 
considerably steeper than that of the galaxy as a whole. If all dust
radiates at the same temperature, we find $T_{\rm d} > 50 K$, at least
twice as high as the whole galaxy under the same assumption. A similar
temperature ($T_{\rm k} >$ 40 K was derived by Sakamoto et al. (1997) 
for the molecular gas in the nuclear disk.

\section{Conclusions}

We have mapped the submillimetre emission from the edge-on galaxy NGC~891
at wavelengths of 850$\mu$m and 450$\mu$m. The appearance of both maps is
very similar to those obtained by others at 1300$\mu$m and in the $^{12}$CO
line. The integrated infrared spectrum of NGC~891 shortwards of 100$\mu$m is 
dominated by emission from PAH particles and very small dust grains. 
Longwards of 100$\mu$m emission from large dust grains dominates. The 
emission spectrum between 100$\mu$m and 1300$\mu$m can be fitted equally 
well by a single population of grains at a temperature of 20 K, or a 
bimodal population of grains at temperatures of 18 K and 27 K. In either
case, we find a total hydrogen mass of the order of 3$\times 10^{9}$ 
M$_{\odot}$, somewhat less than estimated by Gu\'elin et al. (1993),
but close enough to be consistent. As a consequence, we confirm their
conclusion that the CO-to-H$_{2}$ conversion ratio in NGC~891 is
substantially less (by a factor of 2 -- 4) than the canonical values
frequently used. 

The distribution of the various ISM components as a function of distance 
to the center of the galaxy of these ISM components can be described as 
roughly gaussian, with a strong central peak and several lesser peaks 
in the disk superposed. The central peak originates in a compact 
circumnuclear disk ($R = 15''$ or 0.7 kpc; see Garc\'ia-Burillo et al. 
1992; Scoville et al. 1993). The compact circumnuclear disk is embedded 
in soft X-ray emission and is at a temperature of at least 40 K, and 
probably significantly higher.
The lesser peaks probably mark spiral arms seen tangentially. 
Although the submillimetre
continuum emission from dust and the line emissions from CO and CI
show very similar distributions, their intensity ratios vary across the
galaxy. Towards the central region CO and especially CI are relatively 
weak compared to the dust emission at 850$\mu$m. In the socalled `molecular 
ring' we find the reverse: CO and
CI are relatively strong. At greater distances to the center, the CO to
dust emission ratio decreases, whereas the CI to dust emission ratio
increases. This behaviour may be explained by an `empty' bulge region,
by greater CO and CI surface filling factors in tangential spiral arms, 
and by an increase in photo-dissociation effects at greater radial 
distances due to an increasing paucity of dense molecular clouds.

\end{document}